# APPLICATION OF SPOD TO THE ANALYSIS OF SOUND FIELD OF AEROACOUSTIC SOURCES

**O.P. Bychkov and G.A. Faranosov***

*Central Aerohydrodynamic Institute, Moscow, Russia*

*e-mail: georgefalt@rambler.ru

Spectral proper orthogonal decomposition (SPOD) is proposed for identification of the multipole structure of aeroacoustic noise sources from far-field measurements. The method is verified via tests with point multipoles and validated using experimental data on flow-induced cylinder noise and turbulent jet noise.

## INTRODUCTION

In aeroacoustics, one often has to deal with sound sources of different multipole order. According to one of the general formulations of acoustic analogy [1], mass flow pulsations correspond to monopole radiation, force pulsations – to dipole radiation, and Reynolds stress pulsations – to quadrupole radiation. In some flows, higher order multipoles can also be significant [2]. Thus, the identification of different multipole components of the noise emitted by the source under study can be useful from the point of view of non-invasive diagnostics of the physical nature of the source itself. Such identification is not always possible because in the case of arbitrary spatial distribution of sources of different multipole order the problem of determining their type, position and intensity has no unambiguous solution [3-5]. Nevertheless, in some practically realizable cases, the formulation of such a problem could make sense. For example, for relatively slow subsonic flows, aeroacoustic sources are often compact [3] (in all or selected directions), which simplifies the problem of their investigation. Examples of compact sources are vortex rings [6, 7], subsonic flow interacting with bodies [8-10], turbulent subsonic jets [2, 11-13], and plasma actuators at moderate frequencies [14, 15].

At the same time, when analyzing the multipole structure of acoustic radiation from complex sources, one should distinguish between the noise generation and noise propagation effects. As it is known [15], the solution of the wave equation in an unbounded stationary medium on a spherical surface surrounding a sound source can be represented as an expansion over a complete system of spherical functions $Y_l^n = e^{in\varphi} P_l^{|n|}(\cos\theta)$, where $\varphi$ and $\theta$ are, respectively, the azimuthal and polar angles of the spherical coordinate system $(r, \theta, \varphi)$, $P_l^{|n|}(\cos\theta)$ – associated Legendre polynomials, $l \geq |n|$.

For acoustically compact stationary sources located in the origin of the spherical coordinate system, there is an unambiguous correspondence between its "physical" multipole components, associated with certain noise generation mechanisms, and the spherical functions, i.e. the value of $l$ corresponds to the order the physical multipole: $l = 0$ – monopole radiation containing only the axisymmetric azimuthal mode $n = 0$; $l = 1$ – dipole radiation consisting of two azimuthal components $|n| = 0, 1$; $l = 2$ – quadrupole radiation containing three azimuthal components $|n| = 0, 1, 2$, etc. [9, 16].

This unambiguous relation is broken if the source is displaced relative to the origin [9, 17, 18] (the displacement, of course, must be comparable or larger than the radiated wavelength), rotated [19], in motion [2] (convection effects), and/or radiates into an inhomogeneous moving medium [20, 21] (refraction effects). In this case, the representation of the directivities of the physical multipoles through the spherical functions, although still valid, becomes less clear with strengthening of such effects as displacement, motion, and/or rotation, since the physical multipoles are "distributed" over different spherical harmonics.

In aeroacoustic problems, these translational effects (not including rotation) are often one-dimensional, i.e. there is a direction along which the source can be displaced, can move, and/or along which the medium moves. Usually this direction coincides with the direction of the incoming flow. In this case, if we align the polar axis of the coordinate system with this selected direction so that it passes through the source region, the azimuthal structure of the sound field of physical multipoles will not depend on the translational effects, but only their polar directionality will change. Note that the influence of the source shift on its directivity pattern can always be made negligibly small at a sufficiently large distance from the source, so the main propagation effects that can distort (compared to radiation in a static medium) the properties of physical multipoles are the effects of convection and refraction. Thus, the clarity of the multipole decomposition in terms of its relation to the physical mechanisms of noise generation can be kept by replacing the system $e^{in\varphi} P_l^{|n|}(\cos\theta)$ with the system $e^{in\varphi} f_l^{|n|}(\theta, M_s, M)$, where the function $f_l^{|n|}(\theta, M_s, M)$ describes the polar directivity of the basis physical multipoles taking into account the effects of convection/refraction, the values $M_s$ and $M$ denote the Mach numbers of the source convection and medium motion, respectively, and the limit $f_l^{|n|}(\theta, M_s, M) \to P_l^{|n|}(\cos\theta)$ as $M_s, M \to 0$ holds true. This implies that the directivity of the individual azimuthal components of the sound field, defined with respect to the axis directed along the mean flow and passing through the source region, may reflect the peculiarities of the noise generation mechanisms, if interpreted in the above sense.

This interpretation of the azimuthal components of the sound field of aeroacoustic sources is the basis for the azimuthal decomposition technique (ADT) developed at TsAGI [11, 22]. The directivities of the basis multipoles "deformed" by propagation effects can in some cases be estimated theoretically [2, 20, 21, 23, 24], which allows a detailed comparison of noise source models and measurement results [2, 11, 20, 22, 24]. Thus, the determination of the azimuthal composition of the sound field of aeroacoustic sources is a powerful tool for analyzing the physical mechanisms of noise generation.

Measurements of the azimuthal content of the sound field are typically carried out using azimuthal microphone arrays [11, 22, 25-27], and the number of microphones in each cross section should be at least as large as the number of significant azimuthal components, although in the case of sound fields emitted by axisymmetric (on average) flows, the required number of microphones can be halved [28]. An alternative method is to use the azimuthal decomposition of the correlation function, which requires only a pair of microphones, one of which is fixed and serves as a reference, and the other moves in the azimuthal direction around the source [29, 30]. In this case, it is possible to resolve a larger number of the azimuthal modes than when using several stationary microphones, but the measurement system becomes complicated to implement due to the presence of a moving sensor. Note also that such an approach allows reconstructing only the averaged energies of the azimuthal modes (the phase characteristics of the modes are not resolved).

In some cases, due to limitations of the experimental setup and/or measurement system, none of the above methods of azimuthal decomposition can be realized. Such limitations may arise in laboratory conditions due to various geometric constraints (equipment cluttering, proximity of reflecting surfaces, etc.), and, for example, in acoustic measurements of full-scale objects at large test rigs [28], when far-field noise can be measured only at a given azimuthal angle or in a narrow azimuthal sector. In such cases, the question arises whether it is possible to estimate the multipole structure of the source using a measurement system located in one azimuthal plane (in fact, a conventional array of microphones distributed along the arc of a circle around the source).

Recently, methods of noise source localization based on the use of multi-microphone arrays, including those aimed at identifying the multipole structure of the source, have been actively developed in aeroacoustics [31-34]. Such methods, however, require a priori a certain model of the noise source, which may not always be adequate to the phenomenon under study. Thus, the development of methods that use simple microphone distributions and at the same time allow determining the source structure a posteriori, as a result of measurements, is of interest.

In the present work, the method of spectral proper orthogonal decomposition (SPOD) is proposed for such an estimation. In the last decade, SPOD has taken a firm place among the tools used to analyze turbulent flows, allowing, for example, to identify spatio-temporal coherent structures in such flows, to carry out a detailed comparison of theoretical models with the results of physical or numerical experiments, to build low-order models for complex flows in order to implement active flow control approaches (see, for example, the review [35] and relevant references therein). A detailed description of SPOD is given in [36], where it is also discussed in relation to POD (proper orthogonal decomposition) [37], which is already classical in hydrodynamics and, in turn, derived from the principal component analysis (PCA) proposed by Pearson [38]. For aeroacoustic problems, SPOD is often used to identify coherent structures in the turbulent flow region and, in some cases, to estimate the sound emission from such structures [39, 40]. In the present work, in contrast to the above studies, we propose to construct a variant of SPOD focused on the analysis the far-field radiation structure so that it would be possible, as a result of measurements, to define the multipole order of the source and estimate the intensities of its individual multipole components.

The rest of the paper is structured as follows. Section 1 describes the idea of the method based on SPOD and the evaluation of its performance on model examples. Section 2 presents a description of measurements carried out in the anechoic chamber AC-2 of TsAGI and the results of the application of the developed method to the analysis of the obtained experimental data. The summary of the main results of the work concludes the paper.

# 1. EVALUATION OF SPOD APPLICABILITY TO THE ANALYSIS OF THE SOUND FIELD STRUCTURE OF COMPACT SOURCES

Let us give a brief description of SPOD following [35]. Let $u(x,t)$ be a one-dimensional stationary random field with mean zero. In many problems, especially in those where there is a spatial transfer of disturbances, it is convenient to operate with the spectral representation of this field in the form of $\hat{u}(x,f)$ and study the behaviour of individual spectral harmonics, comparing it with various theoretical models.

SPOD allows to find the optimal approximation of the random variable $\hat{u}(x,f)$ by the system of deterministic orthogonal energy-ranked functions $\psi_j(x,f)$ so that

$$\hat{u}(x,f) = \sum_{j=1}^{\infty} a_j(f)\psi_j(x,f).  \qquad (1)$$

The optimality is understood in the sense that at taking $N$ first terms of the expansion (1) the accuracy of the approximation of $\hat{u}(x,f)$ for any other set of basis functions will be lower. To formulate and solve the problem, the inner product is introduced

$$\langle u,v \rangle_{x,t} = \iint v^*(x,t)W(x)u(x,t)dxdt \qquad (2)$$

defining the notion of norm and orthogonality of the considered functions. Here * denotes complex conjugation (and transposition – for matrices), $W(x)$ is a non-negative weight function chosen from physical considerations, the example of which will be shown below for the considered problem. For stationary random fields, all statements about optimality, best approximation, etc. are understood "on average", i.e. in the sense of the mathematical expectation of the corresponding random variables.

It can be shown [36, 37] that in the frequency domain the problem of determining the functions $\psi_j(x,f)$ reduces to the following eigenvalue problem

$$\int S(x,x',f)W(x')\psi(x',f)dx' = \lambda(f)\psi(x,f), \qquad (3)$$

where $S(x,x',f) = \int C(x,x',\tau)e^{-i2\pi f\tau}d\tau$ is the Fourier transform of the cross-correlation tensor $C(x,x',t-t') = E\{u(x,t)u^*(x',t')\}$, and $E\{g\}$ denotes expectation of the random variable g and the hypothesis of stationarity of the considered random processes is used. It can be shown that, for problem (3), there exists a countable set of eigenmodes, i.e. $\{\psi_j, \lambda_j\}$ pairs, which can be ranked by the magnitude of (non-negative) eigenvalues $\lambda_1 \geq \lambda_2 \geq ... \geq 0$. The eigenfunctions $\psi_j(x,f)$ – SPOD modes – are mutually orthogonal in the sense of the spatial inner product

$$\langle \psi_j, \psi_k \rangle_x = \int \psi_k^*(x,f)W(x)\psi_j(x,f)dx = \delta_{jk}, \qquad (4)$$

where $\delta_{jk}$ is the Kronecker delta. In addition, all basis harmonics $\psi_j(x,f)e^{i2\pi ft}$ are mutually orthogonal also in the sense of the original space-time inner product (2). Thus, the spectrum $\hat{u}(x, f)$ of each individual realization of the considered random field can be optimally represented in the form (1), where the coefficients $a_j = \langle \hat{u}, \psi_j \rangle_x$ are the projections of the original field onto the basis functions, and these coefficients are mutually uncorrelated so that $E\{a_j a_k^*\} = \lambda_j \delta_{jk}$. Hence, the cross-spectral density $S(x, x', f)$ is represented in a diagonal form

$$S(x, x', f) = \sum_{j=1}^{\infty} \lambda_j(f) \psi_j(x, f) \psi_j^*(x', f). \tag{5}$$

Such a representation of a stochastic field is useful if only a small number of modes contribute to (5). This allows constructing low-order models of the original phenomenon, as well as to try to relate specific physical mechanisms to individual modes.

Let us further consider how the above method can be adapted to estimate the multipole structure of an acoustically compact noise source using a measurement system located in the fixed azimuthal plane.

Consider a set of uncorrelated multipoles of order $l$ located at the origin. Then the far sound field can be represented as

$$p(r, \theta, \varphi, t) \approx \frac{1}{r} \sum_{n=0}^{l} \left( a_n(t - r/c) \cos n\varphi + b_n(t - r/c) \sin n\varphi \right) P_l^n(\cos \theta), \tag{6}$$

where $a_n$ and $b_n$ are mutually uncorrelated. In the frequency domain Eq. (6) takes the form

$$\hat{p}(r, \theta, \varphi, f) \approx \frac{e^{-i2\pi f r/c}}{r} \sum_{n=0}^{l} \left( \hat{a}_n(f) \cos n\varphi + \hat{b}_n(f) \sin n\varphi \right) P_l^n(\cos \theta).$$

For observation points located in the azimuthal plane $\varphi = const$, without loss of generality we can write

$$\hat{p}(r, \theta, 0, f) \equiv \tilde{p}(r, \theta, f) \approx \frac{e^{-i2\pi f r/c}}{r} \sum_{n=0}^{l} \hat{a}_n(f) P_l^n(\cos \theta). \tag{7}$$

The cross-spectral density for any two points in this plane in the far field, taking into account mutual uncorrelation of separate multipoles, will then be expressed as

$$\hat{S}(r, \theta, r', \theta', f) \approx \sum_{n=0}^{l} A_n(f) P_l^n(\cos \theta) \frac{e^{-i2\pi f r/c}}{r} P_l^n(\cos \theta') \frac{e^{i2\pi f r'/c}}{r'}, \tag{8}$$

where $A_n$ corresponds to the intensity of the sources, i.e. $E\{\hat{a}_n\hat{a}_m^*\} = A_n\delta_{nm}$. One can note the similarity of expressions (8) and (5). Let us consider how we can set the problem of SPOD for the sound field in the given azimuthal plane so that the SPOD modes are close to the functions $P_l^n(\cos\theta)e^{-i2\pi f r/c}/r$ characterizing the noise source. For simplicity, let us assume that the observation points lie on the arc of raduis $r = R$ as is usually the case in acoustic experiments. In this case (8) reduces to

$$S(\theta, \theta', f) \approx \sum_{n=0}^{l} A_n(f) P_l^n(\cos\theta) P_l^n(\cos\theta') . \qquad (9)$$

According to (4), (5) the functions $P_l^n(\cos\theta)$ will be SPOD-modes if the following orthogonality condition is fulfilled for them

$$\left\langle P_l^n, P_l^m \right\rangle_\theta = \int_0^\pi P_l^m(\cos\theta) W(\theta) P_l^n(\cos\theta) d\theta = \delta_{nm} . \qquad (10)$$

As it is known, associated Legendre polynomials are not mutually orthogonal in general, but only some subsets of them are orthogonal [40]. In the problem under consideration, we are interested in the possibility of identifying individual components of multipoles of a given degree $l$. For such a subset $l = const$, the orthogonality has the form

$$\int_{-1}^{1} P_l^m(q) P_l^n(q) \frac{dq}{1-q^2} = \begin{cases} \delta_{mn} \dfrac{(l+n)!}{(l-n)!n}, & n > 0, \\ \infty, & n = 0, \end{cases} \qquad (11)$$

where $q = \cos\theta$ and it is formally shown that the corresponding integral diverges for $n = 0$. In a strict sense, the divergent integral could be a problem, but we will further focus on cases realized in physical or numerical experiments in which we typically deal with a discrete set of coordinate points, in our case – the observation angles $\{\theta_i\}$, $i = 1,...,N$. Let us assume that the angles are uniformly distributed with the step $\Delta\theta$ in the sector $0 < \theta_1 < ... < \theta_N < \pi$. Then condition (11) can be written in the form

$$\sum_{i=1}^{N} P_l^m(\cos\theta_i) P_l^n(\cos\theta_i) \frac{\Delta\theta}{\sin\theta_i} \approx \delta_{mn} g_{ln}, \qquad (12)$$

where $g_{ln} > 0$ – is a set of positive numbers, which are close to $(l+n)!/((l-n)!n)$ at $n > 0$ and close to $\int_{\cos\theta_N}^{\cos\theta_1} \left(P_l^0(q)\right)^2 \left(1-q^2\right)^{-1} dq$ at $n = 0$ provided that the partition of the interval $[0, \pi]$ is sufficiently fine. If we accept the orthogonality condition in the sense of (12), i.e., define the weight function as

$$W(\theta) = \frac{\Delta\theta}{\sin\theta}, \qquad (13)$$

then the corresponding SPOD modes, orthogonal in the sense of the so constructed scalar product

$$\left\langle \psi_j, \psi_k \right\rangle_\theta = \sum_{i=1}^{N} \psi_k^*(\theta_i, f) W(\theta_i) \psi_j(\theta_i, f) = \delta_{jk}, \qquad (14)$$

will be close to the corresponding associated Legendre polynomials (with accuracy up to factors $g_{ln}$ and the error associated with the discretization in $\theta$), and the spectral density of the acoustic radiation in each direction

$$S(\theta, \theta, f) = \sum_{j=1}^{N} \lambda_j(f) |\psi_j(\theta, f)|^2 \qquad (15)$$

will thus be decomposed into energy-ranked functions basis multipole components. Note that it does not matter whether the coefficients $g_{ln}$ are assigned to an eigenvalue or to an eigenfunction, since the emission intensity of each multipole is characterised by the product $\lambda_j(f) |\psi_j(\theta, f)|^2$. The number of significant eigenvalues in such a decomposition must be equal to $l+1$ – the number of basis multipoles of the given degree contributing to the radiation in the fixed azimuthal plane.

The problem (3) in discrete form can be written for each frequency component $f$ as [35]

$$\mathbf{S}_f \mathbf{W} \mathbf{\Psi}_f = \mathbf{\Psi}_f \mathbf{\Lambda}_f, \qquad (16)$$

where all matrices have the size $N \times N$ and consist of the following elements: $\mathbf{\Psi}_{f\,ij} = \psi_j(\theta_i, f)$, $\mathbf{\Lambda}_{f\,ij} = \lambda_j(f) \delta_{ij}$, $\mathbf{W}_{ij} = \Delta\theta/\sin\theta_i \, \delta_{ij}$, $\mathbf{S}_{f\,ij}$ – elements of the cross-spectral matrix for frequency $f$. To determine $\mathbf{S}_{f\,ij}$, the standard Welch's method [41] is used in the form proposed in [35] as briefly described below.

Let the total length of the time history of the random process recorded at the observation positions with time step $\Delta t$ contains $M$ points. It is convenient to form the corresponding set of signals for $N$ observation points into a matrix $\mathbf{Q} = [\mathbf{q}_1,...,\mathbf{q}_M]$ of size $N \times M$. The rows of $\mathbf{Q}$ represent time histories at the corresponding observation points, and the columns – snapshots of the stochastic field on the set of observation points. Next, the full time history is partitioned into $N_b$ blocks of length $N_f$ each, and for each $n$th block a matrix $\mathbf{Q}^{(n)} = [\mathbf{q}_1^{(n)},...,\mathbf{q}_{N_f}^{(n)}]$, of size $N \times N_f$, is constructed from the corresponding columns of matrix $\mathbf{Q}$. Then, matrices $\mathbf{Q}^{(n)}$ are converted into frequency space $\mathbf{Q}^{(n)} \to \hat{\mathbf{Q}}^{(n)} = [\hat{\mathbf{q}}_1^{(n)},...,\hat{\mathbf{q}}_{N_f}^{(n)}]$ by applying a discrete Fourier transform $\hat{\mathbf{q}}_k^{(n)} = N_f^{-1/2} \sum_{j=1}^{N_f} \mathbf{q}_j^{(n)} \exp(-i2\pi f_k(j-1)\Delta t)$, where $f_k = (k-1)/(N_f \Delta t)$, $k = 1,...,N_f$ (if necessary, the transform can be performed with a window function). Thus, the cross-spectral matrix is estimated by averaging over $N_b$ blocks, and for a given frequency $f_k$ is represented by $\mathbf{S}_{f_k} = \hat{\mathbf{Q}}_{f_k} \hat{\mathbf{Q}}_{f_k}^*$, where $\hat{\mathbf{Q}}_{f_k} = \sqrt{\Delta t / N_b} [\mathbf{q}_k^{(1)},...,\mathbf{q}_k^{(N_b)}]$ is a matrix consisting of $k$th columns of matrices $\mathbf{Q}^{(n)}$.

Since $\tilde{\mathbf{S}}_{f_k} = \mathbf{S}_{f_k} \mathbf{W}$ is the Hermitian matrix, an eigen decomposition $\tilde{\mathbf{S}}_{f_k} = \mathbf{U}_{f_k} \mathbf{\Lambda}_{f_k} \mathbf{U}_{f_k}^*$ with orthonormal basis vectors exists. This decomposition can be performed using standard tools. Therefore, SPOD solution is

$$\mathbf{\Psi}_{f_k} = \mathbf{W}^{-1/2} \mathbf{U}_{f_k}. \tag{17}$$

These SPOD modes are orthogonal in the sense of the defined inner product $\mathbf{\Psi}_{f_k}^* \mathbf{W} \mathbf{\Psi}_{f_k} = \mathbf{I}$ (it is a matrix form of (14)). For the cross-spectral density, one then has the following decomposition

$$\mathbf{S}_{f_k} = \mathbf{\Psi}_{f_k} \mathbf{\Lambda}_{f_k} \mathbf{\Psi}_{f_k}^*, \tag{18}$$

which is a matrix form of (15). In a practical sense, relation (18) means that the acoustic radiation intensity at a given frequency $f_k$ (diagonal elements of the matrix $\mathbf{S}_{f_k}$), obtained in a discrete set of spatial points $\theta_i$, is represented as a superposition of $N$ modes

$$\mathbf{s}_{f_k} = \sum_{j=1}^{N} \lambda_{f_k j} | \mathbf{\psi}_{f_k j} |^2, \tag{19}$$

where $\mathbf{s}_{f_k}$ is a vector composed of diagonal elements of matrix $\mathbf{S}_{f_k}$, $\lambda_{f_k j}$ are the eigenvalues – diagonal elements of matrix $\mathbf{\Lambda}_{f_k}$, $\mathbf{\psi}_{f_k j}$ are the corresponding eigenvectors – columns of matrix $\mathbf{\Psi}_{f_k}$. According to the constructed procedure, the first several terms in (19) determine the directivities and intensities of the basis multipole sources. As a measure of the intensity of each multipole one can take the value at the maximum of its directivity pattern

$$w_{f_k j} = \lambda_{f_k j} \max_j [\mathbf{\psi}^*_{f_k j} \mathbf{\psi}_{f_k j}]. \tag{20}$$

If necessary, after determining the connection between eigenvectors and azimuthal modes the total acoustic power for each multipole can be calculated.

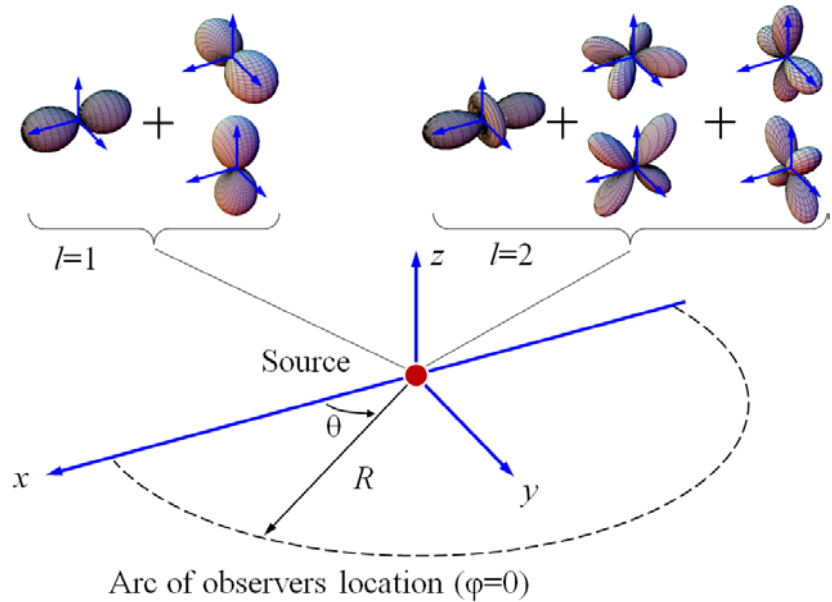

Fig. 1. Model sources consisting of uncorrelated dipoles ($l=1$) or quadrupoles ($l=2$).

Let us consider several model cases with parameters typical for laboratory aeroacoustic experiments. Suppose there are three basic dipole sources at the origin, i.e., with a multipole order of $l=1$, with directivity patterns characterized by the functions $P_1^0(\cos\theta) = \cos\theta$, $P_1^1(\cos\theta)\cos\varphi = \sin\theta\cos\varphi$ and $P_1^1(\cos\theta)\sin\varphi = \sin\theta\sin\varphi$, with equal amplitudes. Let the sources be uncorrelated and radiate sound at a frequency of $f_0 = 1\text{kHz}$ (see Fig. 1). In this case, uncorrelation means that in each finite realization, the initial phase of each source is a random variable uniformly distributed over an interval $[0, 2\pi)$, and the mean value of the other relevant random variables is calculated by averaging over multiple realizations of the stochastic process.

Consider observation points in the far field that lie on the arc of radius $R = 1$ m in a plane defined by the azimuthal angle $\varphi = 0$ (see Fig. 1). At these locations, the received signal in each realization of the process will have the form

$$\hat{p}_{f_0}(R,\theta,0,t) \approx \frac{e^{i2\pi f_0(t-R/c)}}{R}(a_0 e^{-i\alpha_0}\cos\theta + a_1 e^{-i\alpha_1}\sin\theta), \qquad (21)$$

where $a_0$, $a_1$ – amplitudes of the sources, $\alpha_0$, $\alpha_1$ – initial phases of the sources (randomly different in different realizations).

Another model source can be created using quadrupoles, whose radiation can be described by an expression

$$\hat{p}_{f_0}(R,\theta,0,t) \approx \frac{e^{i2\pi f_0(t-R/c)}}{R}\left(a_0 e^{-i\alpha_0}\frac{1}{2}(3\cos^2\theta - 1) + a_1 e^{-i\alpha_1}\frac{3}{2}\sin 2\theta + a_2 e^{-i\alpha_2} 3\sin^2\theta\right). (22)$$

In the model examples with dipole sources, three cases with different amplitude ratios were examined: $a_0 = 2a_1$, $a_1 = 2a_0$ and $a_0 = a_1$. Note that the case where the source amplitudes are approximately equal and the overall radiation is nearly isotropic is the most significant for analysis, as otherwise (domination of a single source), the nature of the multipole can be inferred directly from the radiation pattern's shape. The signals were discretely sampled at increments of $\Delta t = 10^{-5}$ s, with a length of each realization $N_f = 1024$, and the number of realizations was $N_b = 100$. The number of observation points $N$ ranged from 10 to 30, which is typical for aeroacoustic experiments.

The results of SPOD algorithm are presented in Fig. 2. Figures 2a-2c show the results for different source amplitudes at $N = 10$. The lines represent the directivity of the initially specified sources, while the symbols indicate the SPOD-modes. It can be observed that in all cases, there is a strong correspondence between the SPOD-modes and the original sources in terms of amplitude and directivity. The number of significant SPOD-modes corresponds to the number of multipoles of the given order, the modes being energy-ranked. An increase in the number of observation points improves the accuracy of the source reconstruction (Fig. 2d).

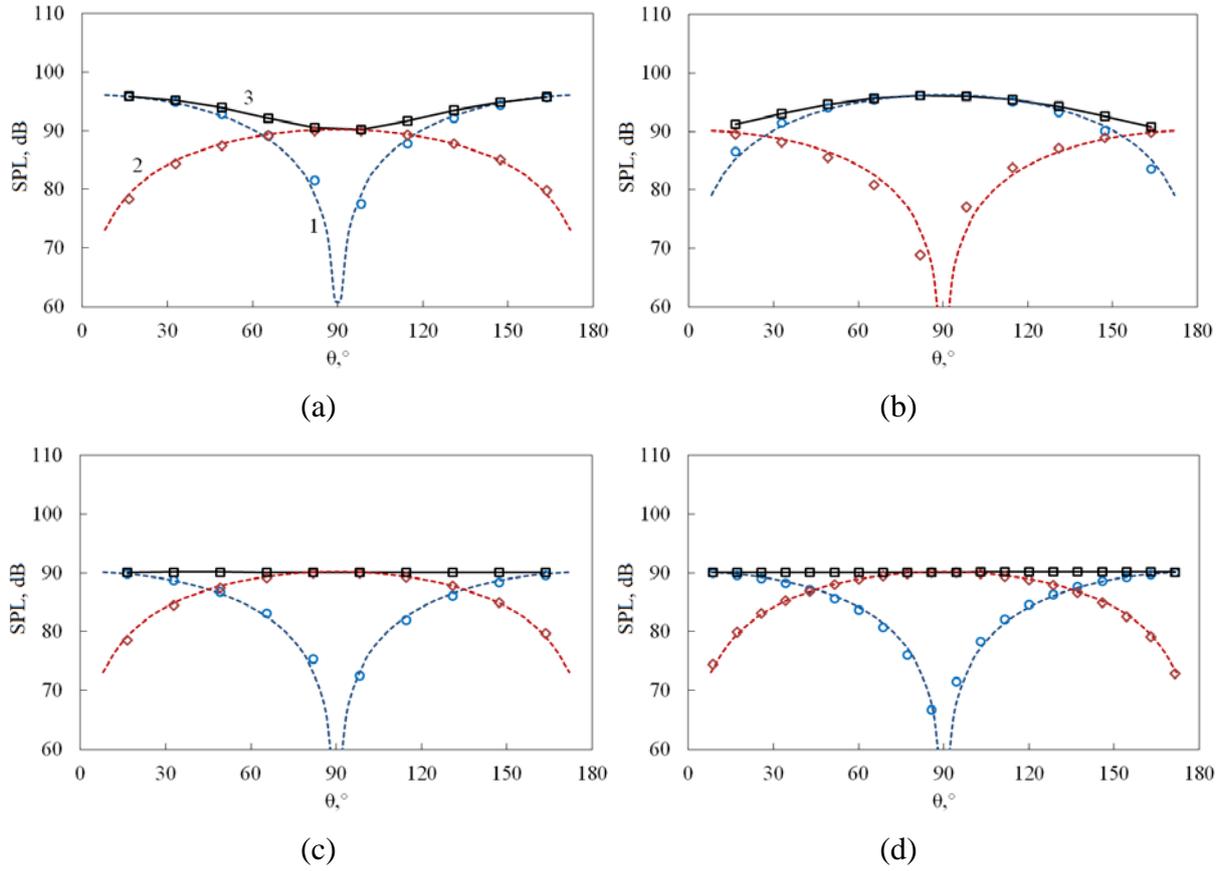

Fig. 2. Directivity patterns of the initial dipole sources (lines) and SPOD-modes (symbols). 1 – axisymmetric dipole ($n=0$); 2 – transverse dipole ($n=1$); 3 – total noise. ○ – SPOD-mode $j=1$; ◊ – SPOD-mode $j=2$; □ – total intensity of SPOD-modes. (a) – $a_0=2$, $a_1=1$, $N=10$; (b) – $a_0=1$, $a_1=2$, $N=10$; (c) – $a_0=1$, $a_1=1$, $N=10$; (d) – $a_0=1$, $a_1=1$, $N=20$.

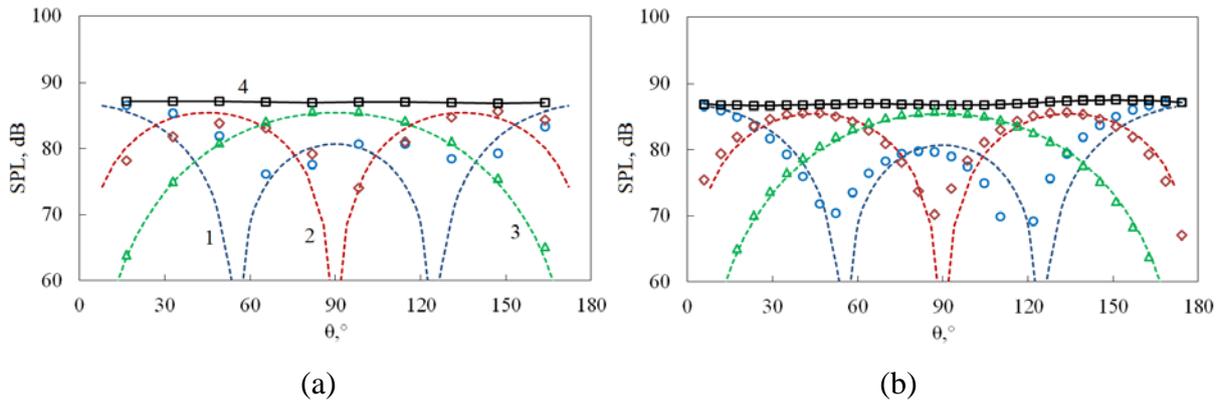

Fig. 3. Directivity patterns of the initial quadrupole sources (lines) and SPOD-modes (symbols) for $N=10$ (a) and $N=30$ (b). 1 – axisymmetric quadrupole ($n=0$); 2 – quadrupole ($n=1$); 3 – quadrupole ($n=2$); 4 – total noise. ○ – SPOD-mode $j=1$; ◊ – SPOD-mode $j=2$; Δ – SPOD-мода $j=3$; □ – total intensity of SPOD-modes.

For quadrupole sources, we considered the case where their total radiation is isotropic on average, which occurs when $a_1 = 2a_2 = a_0/(3\sqrt{3})$, and $a_0 = 2$ was chosen for calculations. The results of SPOD are shown in Fig. 3 for the cases with $N=10$ and $N=30$ observation points. As can be seen, with a smaller number of observation points, the amplitudes of the different modes are recovered quite accurately. However, the directivities of the SPOD-modes in regions far from the maxima may deviate from those of the original sources. This is because as the order of multipole increases, and therefore the number of lobes in its directivity pattern increases, the orthogonality condition in a discrete form written for a fixed number of observation points $N$ will approximate the original condition worse and worse. Therefore, the shapes of the SPOD-modes that satisfy this condition exactly will increasingly deviate from the directivities of the original sources. This problem is solved by increasing the number of spatial grid points (Fig. 3b).

## 2. VALIDATION OF THE METHOD

To experimentally validate the developed algorithm, flow-induced cylinder noise measurements were performed in the anechoic chamber AC-2 of TsAGI. As is well known [1, 8-10], dipole-like noise sources are generated in this case, and two orthogonal dipole components appear: the so-called lift dipole and drag dipole.

A steel cylinder of diameter $d = 5$ mm was inserted into a turbulent jet issuing from a circular nozzle of diameter $D = 40$ mm at a velocity $V_j = 100$ m/s. The cylinder was positioned in the turbulent flow region, 250 mm downstream of the nozzle exit (Fig. 4). This configuration corresponded to the one described in [33], which made it possible to compare the source identification results obtained using different methods. Noise was measured using nine Bruel & Kjaer microphones (type 4189) that were evenly distributed in increments of $\Delta\theta=15°$ along the arc of radius $R = 1.08$ m, centered at the point of intersection of the cylinder and the jet axis, as illustrated in Fig. 4. Measurements were conducted for both the free jet and the jet in the presence of the cylinder. The cylinder was positioned such that its axis was perpendicular to the plane containing the microphones (Fig. 4). Under these conditions, the dipole moments associated with fluctuations of the lift and drag forces (indicated by arrows in Fig. 4) lie within the plane of the microphone array, resulting in a sound field that is approximately described by Eq. (18).

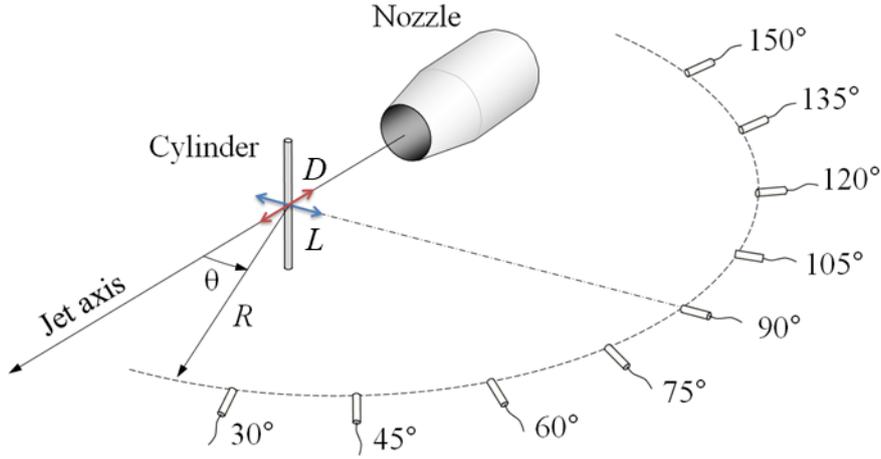

Fig. 4. Sketch of the experimental setup. Arrows show orientations of the dipole components of the flow-induced cylinder noise: the lift dipole (*L*) and the drag dipole (*D*).

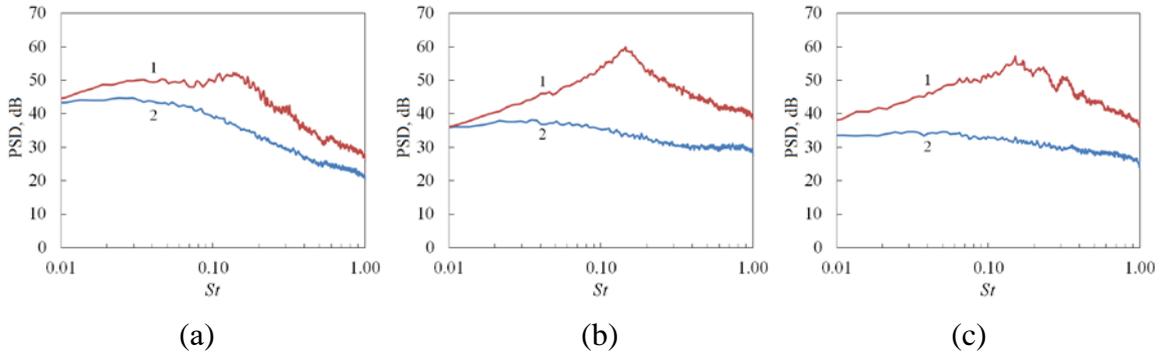

(a)             (b)             (c)

Fig. 5. Measured noise spectra: (a) – $\theta=30°$; (b) – $\theta=90°$; (c) – $\theta=150°$.
1 – cylinder in jet; 2 – free jet.

In the presence of the cylinder, the total noise of the system is dominated by the noise of the cylinder, as shown in Fig. 5, since the jet noise is quadrupole in nature, and in low-velocity flows its intensity is small compared to the intensity of dipole sources [1]. In Fig. 5, the Strouhal number $St = fd/V_j$ is calculated from the cylinder diameter and jet velocity. The maximum noise of the cylinder corresponds to the Strouhal number about $St=0.14$, which corresponds to the results obtained in [33].

Figure 6 shows the spectra of the SPOD eigenvalues for the sound field in the presence of the cylinder in the jet (the modes are energy-ranked, and the number of modes is equal to the number of microphones). As can be seen, the first two SPOD-modes dominate in almost the entire frequency range, which indirectly indicates the dipole nature of the noise. It is interesting to note that at $St\approx0.06$, the intensities of the two modes become equal, and spectra of the first and second modes continuously transform into each other. This effect also indicates that the SPOD-modes defined by the proposed algorithm correspond to physical objects associated with certain noise generation mechanisms.

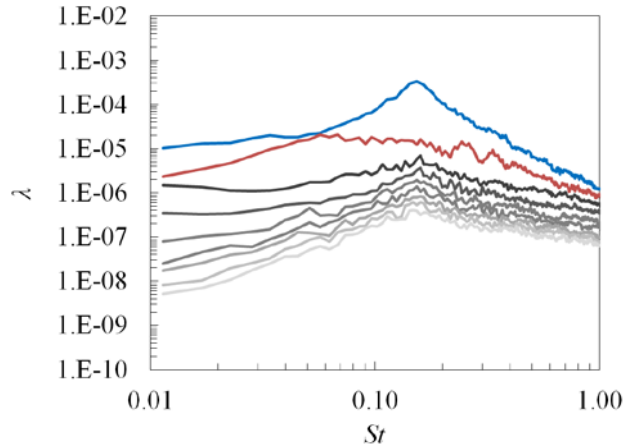

Fig. 6. Spectra of the SPOD eigenvalues for the sound field
for the case of the cylinder in the jet.

The connection of SPOD-modes with physical multipoles is also confirmed by the structure of their directivity, shown in Fig. 7 for different frequency bands. Only two dominant modes are shown. It can be seen that one of them has a single-lobed structure, while the other has a two-lobed structure (and the neighboring lobes are always in antiphase). For the noise generated by the cylinder in the jet, the dominance of dipole noise is expected, and the described features of the SPOD-modes indeed correspond to different components of dipole radiation. In the low-frequency region ($St < 0.06$), the radiation of a longitudinal drag dipole dominates (Fig. 7a), and at frequencies above $St=0.06$, including in the spectral peak region, the transverse lift dipole becomes decisive (Fig. 7b-d). Deviations of the directivity shapes those of the "ideal" dipoles may be associated with the effects of convection and refraction, which "deform" the directivities of the initial basic sources. This deformation may lead to a certain shift in the positions of the minima of the directivity patterns due to the condition of mutual orthogonality of the SPOD modes (Fig. 7a). Another reason may be related with small amplitude of one of the sources, close to the boundary of the dynamic range, as a result of which the characteristics of its directivity are recovered with a noticeable error (Fig. 7c, d). At the same time, it should be noted that the downstream convective amplification of the radiation, which is clearly visible in the total noise (Fig. 7a), is generally quite correctly transferred by the SPOD algorithm to the corresponding dipole components, at least in the area of the directivity maxima.

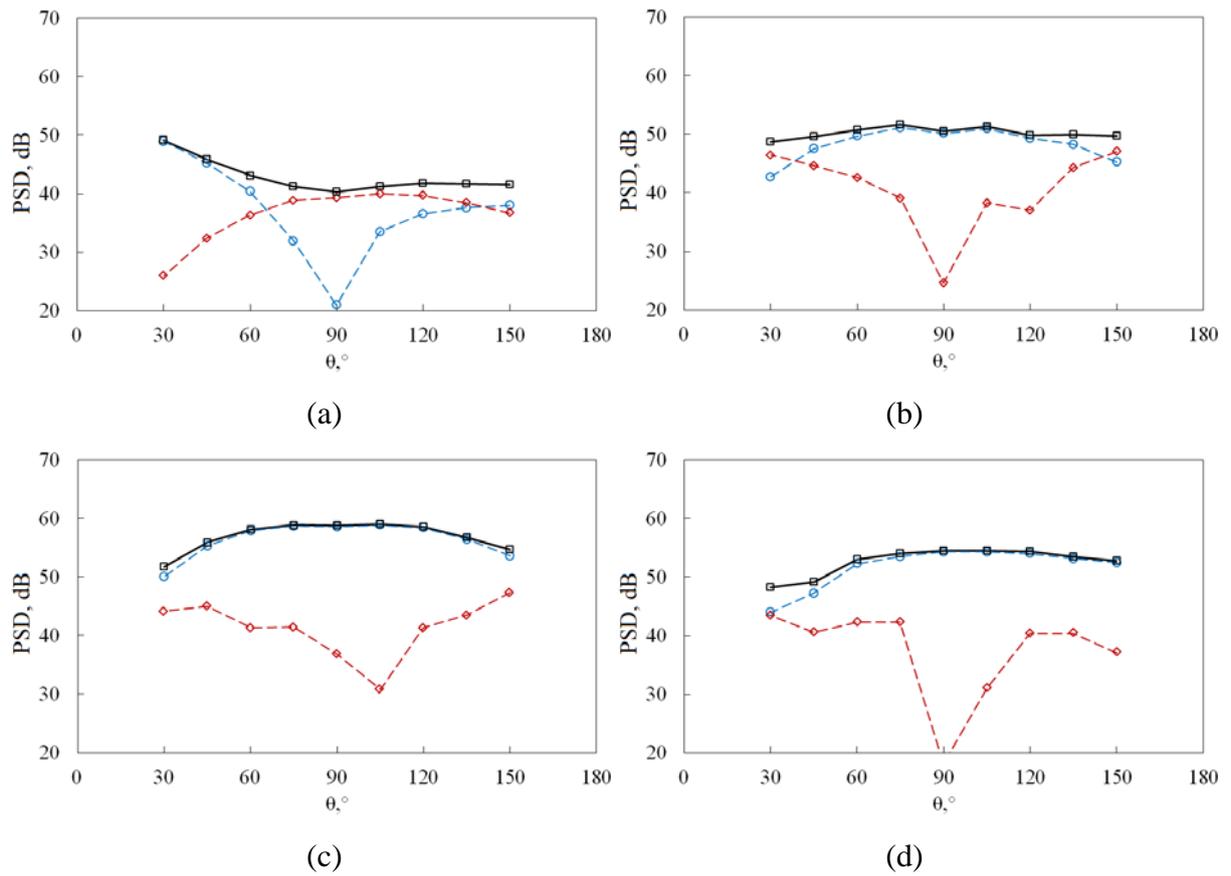

Fig. 7. Directivity patterns at different frequencies for the case of the cylinder in the jet: (a) – $St$=0.017; (b) – 0.08; (c) – 0.14; (d) – 0.2. Solid line – measurements; ○ – SPOD-mode $j=1$; ◊ – SPOD-mode $j=2$; □ – total intensity of SPOD-modes.

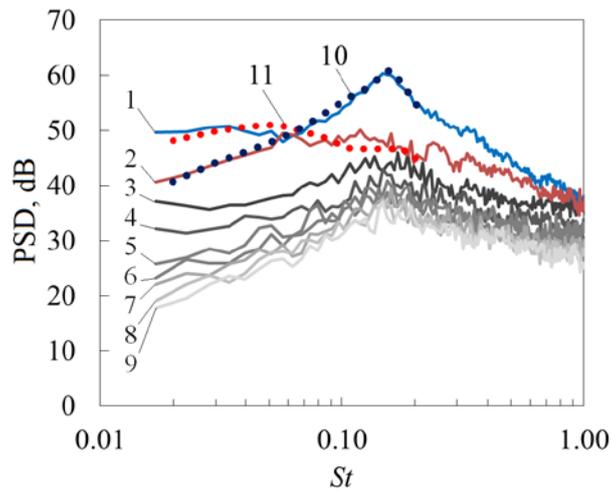

Fig. 8. Power spectral densities for the radiation intensity of individual SPOD-modes (1-9), calculated by the formula (20), and power spectral densities of lift (10) and drag (11) dipoles, determined using ADT [33] (data are scaled to a distance of 1 m from the source).

As noted above, the value in the maximum of directivity pattern of each mode can serve as a quantitative measure of the radiation intensity (for SPOD modes, this definition corresponds to formula (20)). A similar value was chosen in [33], which allows for a direct comparison of the spectral characteristics of dipole sources obtained by ADT in [33] and SPOD as proposed in this work. This comparison is shown in Fig. 8, and it demonstrates the coincidence of the results of estimating the intensities of the dipole components determined by various methods. In this case, the SPOD-based approach uses a one-dimensional array of microphones located in the fixed azimuthal plane, which in some cases makes it easier to implement compared to ADT.

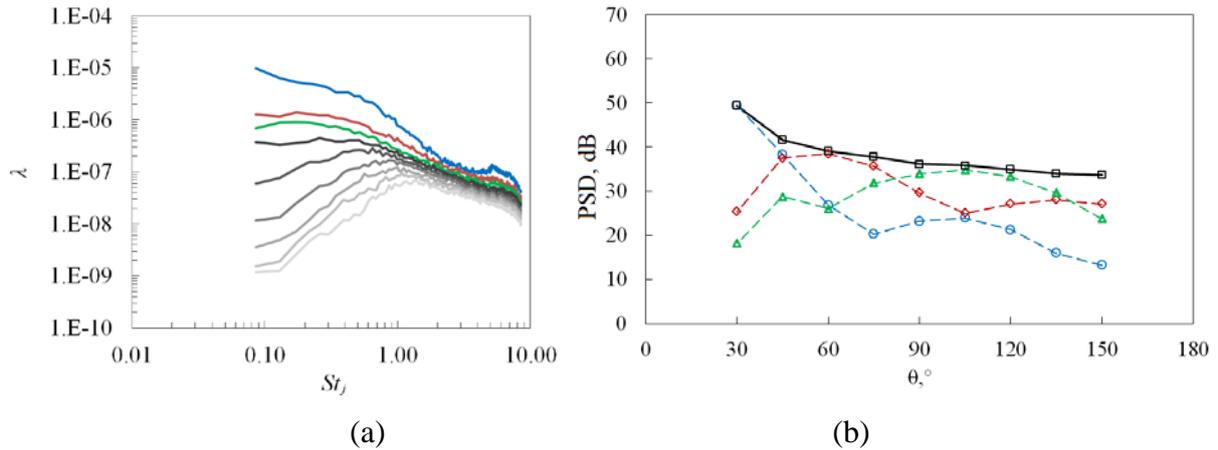

(a) (b)

Fig. 9. (a) – Spectra of SPOD eigenvalues for the sound field of the free jet; (b) – directivity patterns of the dominant SPOD-modes for $St_j$=0.1. See Fig. 3 for the notation.

Figure 9 shows the results of SPOD for the sound field of the free jet. Note that for the jet noise analysis, as is usually accepted, we use the Strouhal number $St_j = fD/V_j$ calculated from the jet velocity and nozzle diameter. The jet velocity is rather small so that the maximum of jet noise spectrum falls in the frequency range of 300-600 Hz, which corresponds to $St_j$=0.1...0.2. In the region of maximum radiation, as can be seen from the spectra of the SPOD eigenvalues shown in Fig. 9a, three modes contribute to the sound field, which indicates the quadrupole nature of the radiation [11]. Note that the SPOD-based approach proposed in this paper assumes that the sound source is compact and located in the center of the arc on which the microphones are located. At the same time, as is known [3], location of noise sources in turbulent jets depends on their frequency, so that higher-frequency sources are located closer to the nozzle. Therefore, the directivities of individual SPOD-modes will be close to those of the physical multipoles for the frequency corresponding to the sources located in the area of the microphone arc center, which, in the considered experiment, was at 6.25$D$ downstream from the nozzle exit. For this distance from the nozzle, the characteristic radiation frequencies correspond exactly to the spectral maximum, i.e. to the frequency range $St_j$=0.1...0.2 [3, 24]. At the same time, since jet noise sources, even for a narrow frequency range, are not compact in the longitudinal direction [24], the correspondence between the directivities of the SPOD-modes and those of point quadrupoles will not be completely accurate. This is seen from Fig. 9b, which shows the directivity patterns of the dominant SPOD-modes for $St_j$=0.1. One can see that shapes of SPOD-modes qualitatively correspond to quadrupoles of different azimuthal order, the shapes of which can be obtained with high accuracy using ADT [11, 27]. However, there is some difference in the location of the minima associated with convective/refractive noise amplification downstream [20] (discussed above in relation to cylinder noise), as well as in the amplitudes of those lobes that are at the boundary of the dynamic range of the method (~10 dB below the dominant mode), which is partly due to the non-compactness of the sources. Nevertheless, the general structure of SPOD-modes allows them to be quite definitely attributed to the corresponding quadrupole sources, i.e., to adequately assess the dominant mechanism of noise generation by the studied flow.

## CONCLUSION

It is proposed to use the method of spectral proper orthogonal decomposition (SPOD) to identify the multipole structure of aeroacoustic sources from far-field measurements. The advantage of SPOD is that it can be implemented within the framework of a simple microphone array located in one azimuthal plane, while the azimuthal decomposition technique (ADT), with which such a problem can be solved most accurately, requires the use of azimuthal microphone arrays covering a significant proportion of the solid angle of noise radiation. It is shown that, given the compactness of the source of a certain (unknown a priory) multipole order, and its location in the center of a circular microphone array covering a sector close to 180°, it is possible, by proper definition of an inner product, to construct such a SPOD procedure in which the dominant SPOD-modes correspond to the physical multipole components of the radiated noise.

The method is tested on several model examples with point multipoles, and also on the experimental measurements of flow-induced cylinder noise and jet noise. The effects of convection and refraction reduce the accuracy of reconstructing the directivities of individual multipoles. However, for low Mach number flows, the accuracy of the intensity estimation for the dominant multipole components remains acceptable. In the presence of theoretical models of the effects of convection and refraction on the directivity pattern of basic multipoles, the definition of the inner product can be further adjusted, thereby increasing the accuracy of the SPOD reconstruction of the multipole structure.

In conclusion, we note the difference between the developed SPOD-based approach and the beamforming method used, for example, in [31-33] to identify dipole noise sources. Both approaches are related to the analysis of the cross-spectral matrix measured by an array of microphones. In beamforming-type methods, such a matrix is modeled using initially selected source models of a given multipole order. In this approach, the correctness of the result depends on the adequacy of the set of sources embedded in the algorithm. Its advantage is the ability to determine the location of noise sources (unknown a priori), the disadvantage is the need to predefine the source type. In the proposed SPOD-based method, cross-spectral matrix is decomposed over an empirical orthogonal basis, which always turns out to be possible, and under certain conditions, the components of this basis can correspond to physically interpretable multipoles. The advantage of this method is the ability to determine the order of multipole of the source as a result of the measurements, the disadvantage is the need for a priori knowledge of the source position.

Thus, the proposed SPOD-based approach, beamforming, as well as ADT are complementary tools and can be used separately or simultaneously for the analysis of complex aeroacoustic systems.

ACKNOWLEDGMENTS

The experimental part of the study was carried out in TsAGI AC-2 anechoic chamber with flow, upgraded with the support of the Ministry of Science and Higher Education of the Russian Federation.

FUNDING

The study was financially supported by the Russian Science Foundation (grant no. 19-71-10064).